\documentclass[12pt,a4paper]{article}
\usepackage[sc,noBBpl]{mathpazo}
\usepackage[mathscr]{eucal}
\usepackage{amsmath,amsfonts,amssymb,amsthm}
\usepackage{graphicx}
\usepackage{mathtools}
\usepackage{color}
\usepackage{tgpagella}
\usepackage{subfigure}
\usepackage{caption}
\usepackage{wrapfig}
\usepackage{float}
\newcommand\scN{{\mathscr N}}

\newcommand\mvector{\boldsymbol}

\newcommand\vc{\mvector{c}}

\newcommand\ve{\mvector{e}}

\newcommand\vp{\mvector{p}}

\newcommand\vr{\mvector{r}}

\newcommand\vZ{\mvector{Z}}
\newcommand\field{\mathbb}

\newcommand\Z{\field{Z}}

\newcommand\Dz{\frac{\mathrm{d}\phantom{z} }{ \mathrm{d}z}}

\usepackage{url}

\theoremstyle{plain}
\newtheorem{theorem}{Theorem}

\newtheoremstyle{note}{\topsep}{\topsep}{\slshape}{}{\scshape}{}{ }{}
\theoremstyle{note}

\numberwithin{equation}{section}
\numberwithin{theorem}{section}
\numberwithin{definition}{section}
\numberwithin{lemma}{section}
\numberwithin{proposition}{section}
\numberwithin{corollary}{section}
\numberwithin{remark}{section}

\mathtoolsset{mathic,centercolon}

\usepackage{subfigure}

\title{Constrained N-body problems}
\author{
   Wojciech Szumi\'nski and Maria Przybylska \\[1em]
  {}Institute of Physics, \\ University of Zielona G\'ora, 
  Licealna 9,  \\
  PL-65-407,  Zielona G\'ora, Poland}
\begin{document}
%
\maketitle

\begin{abstract}
We consider a problem of mass points
  interacting gravitationally whose motion is subjected to certain
  holonomic constraints. The motion of points is restricted to certain
  curves and surfaces. We illustrate the complicated
  behaviour of  trajectories of these systems using Poincar\'e cross sections. For
  some models we prove the non-integrability analysing properties of
  the differential Galois group of variational equations along certain
  particular solutions of considered systems. Also some integrable
  cases are identified.
\end{abstract}
\date{\small Key words: n-body problem; non-integrability; Morales--Ramis theory; differential  Galois theory; Poincar\'e sections; chaotic Hamiltonian systems.}
\maketitle

\section{Introduction}
\label{sec:intro}
Let us consider several point masses interacting mutually according to
a certain low.  This is just the $n$-body problem.  For the classical
gravitational, or electrostatic interactions such problem with $n>2$
is not integrable.  Let us restrict the motion of points to certain
surfaces or curves. These holonomic constrains modify interactions of
points. In some cases these modifications lead to the non-integrability,
and in others to the integrability.  The described constrained
classical $n$-body problems can be considered as a source of toy models
for testing various methods and tools for study dynamics of
classical systems. In fact this paper arose from such investigations.
Several simple examples show that, in fact, one can meet interesting
and difficult problems investigating this kind of systems and moreover, such,  let us say, academic investigations, give unexpected results.

To describe them let us recall the anisotropic Kepler problem which
appears in quantum mechanics of solid. It was thoroughly investigated
by Guztwiller \cite{Gutzwiller::90}.  The rescaled Hamiltonian of the problem is given by
\begin{equation}
  \label{eq:h3}
  H= \frac{1}{2} \left(p_1^2 +  p_2^2+  p_3^2\right) -\frac{1}{\sqrt{x^2+\mu(y^2+z^2)}},
\end{equation}
where $\mu$ is a positive constant. For the two-degrees of freedom
version of this problem the Hamiltonian reads
\begin{equation}
  \label{eq:h2}
  H= \frac{1}{2}\left( p_1^2+   p_2^2 \right)-\frac{1}{\sqrt{x^2+\mu y^2}}.
\end{equation}

\begin{figure}[htpp]
  \centering \subfigure[Geometry of  model 1; ]{
    \includegraphics[width=0.29\textwidth]{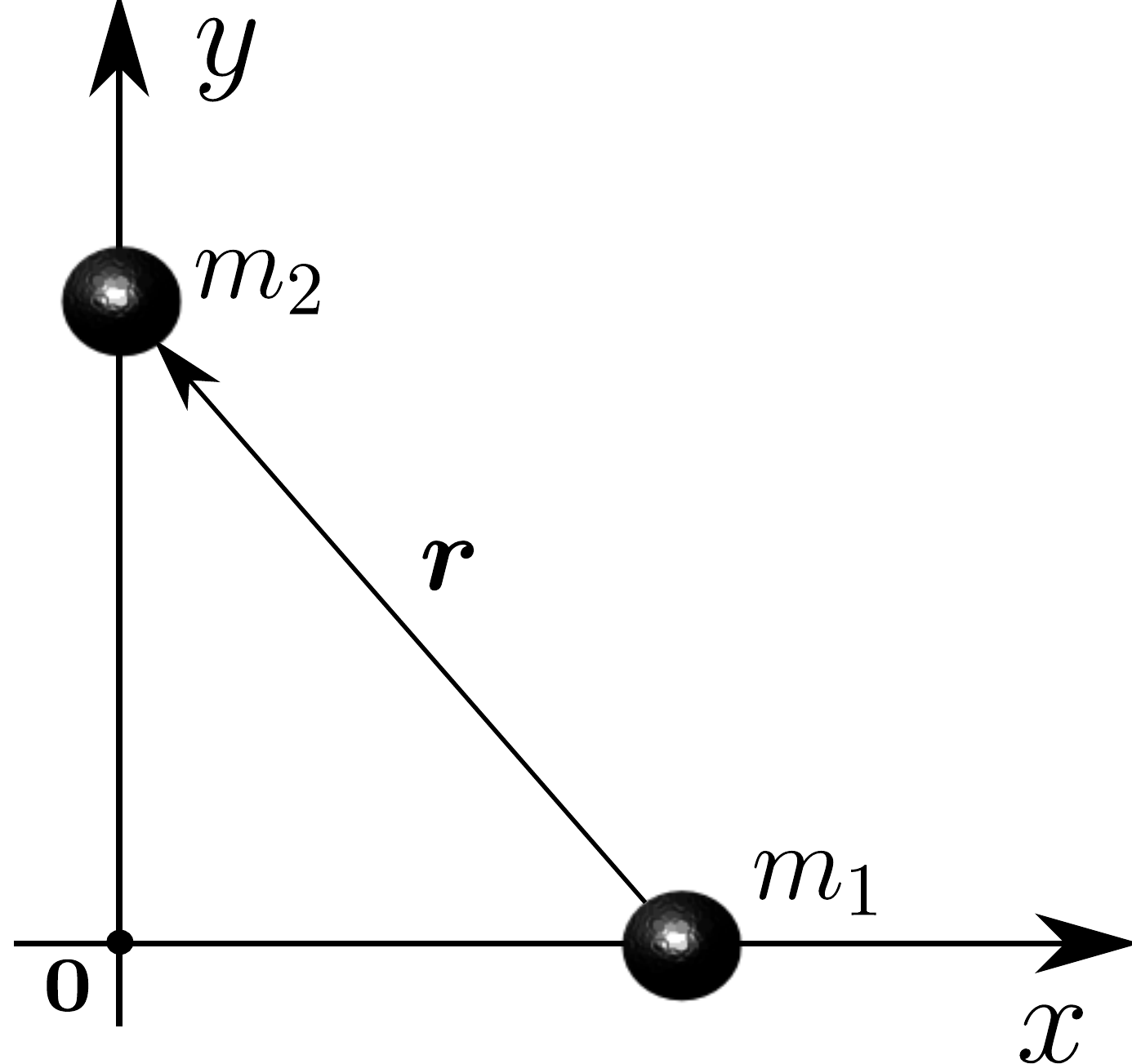}
    \hspace{0.2in} \label{fig:intro(a)} } \ \ \ \ \ \ \ \ \
  \subfigure[Geometry of  model 2.]{
    \includegraphics[width=0.4\textwidth]{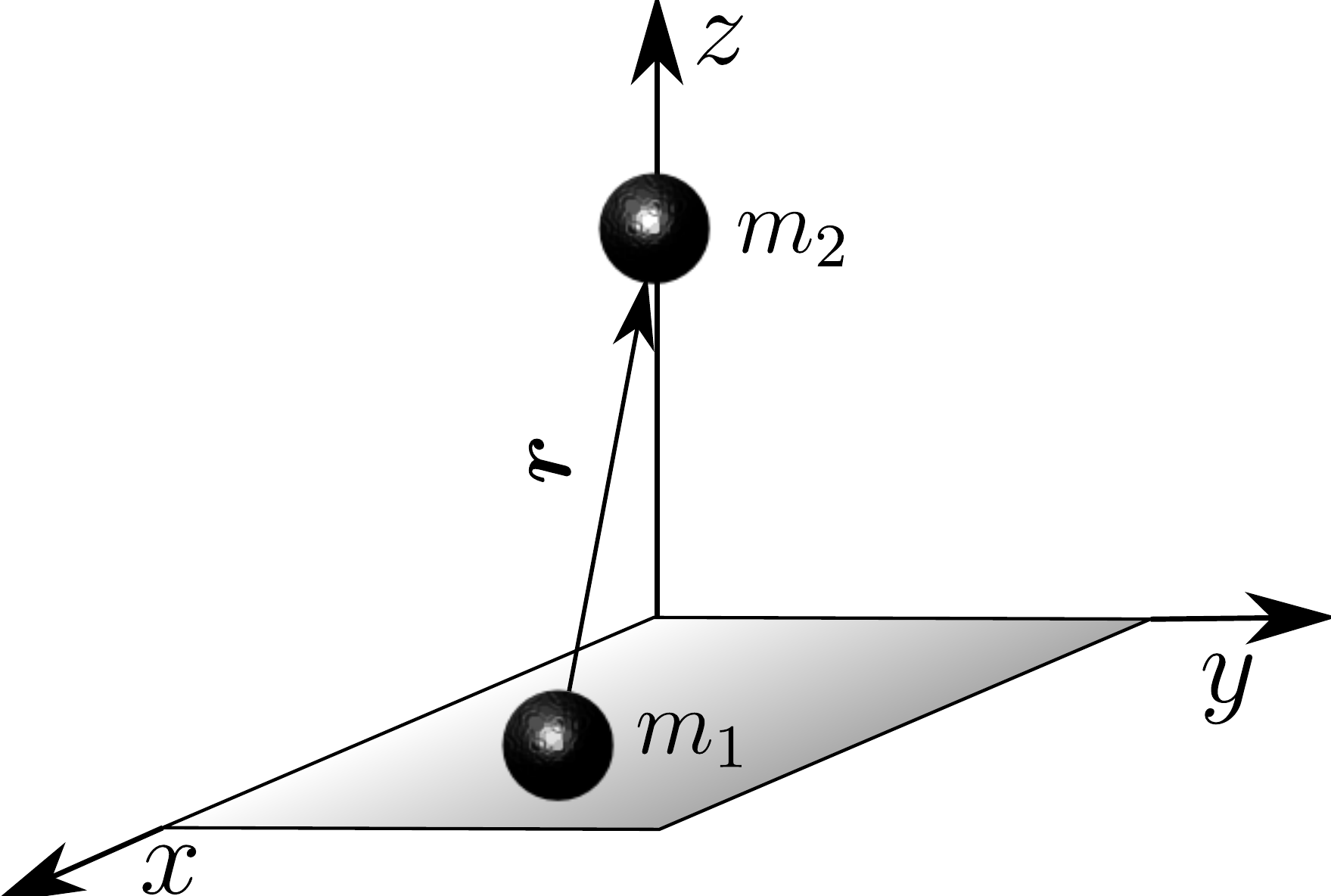}\label{fig:intro(b)}
  }
  \caption{Motion of two masses on: (a) perpendicular axes, (b) plane
    and perpendicular axis.\label{fig:intro}}
\end{figure}

Unexpectedly, these systems can be considered as gravitational two
body problems with constraints. To see this, let us consider two
masses, one mass moving  along a line, and the second mass moving along a
perpendicular line, see Fig.~\ref{fig:intro(a)}. The Hamiltonian of the
system is following
\begin{equation}
  \label{eq:intro}
  H_{1}=\frac{p_1^2}{2m_1}+\frac{p_2^2}{2m_2}-\frac{Gm_1m_2}{\sqrt{x^2+y^2}}.
\end{equation}
So, by a simple rescaling we obtain
Hamiltonian~\eqref{eq:h2}. Similarly, let one mass moves along a line, and
the other moves in a plane perpendicular to this line, see Fig.~\ref{fig:intro(b)}.  The Hamiltonian
has the form
\begin{equation}
  \label{eq:intro1}
  H_{2}=\frac{p_1^2}{2m_1}+\frac{1}{2m_2}(p_2^3+p_3^2)-\frac{Gm_1m_2}{\sqrt{x^2+y^2+z^2}}
\end{equation} 
and again its simple rescaling gives~\eqref{eq:h3}.

As we can see the Hamiltonians \eqref{eq:h2} and \eqref{eq:h3} differ
from the Hamiltonians of standard planar and spatial Kepler problem
only in the parameter $\mu$. For $\mu\neq0$, contrary
to standard Kepler problem, the force is not radial.  The dynamics of anisotropic Kepler problem
is dramatically different from that of the standard Kepler problem.

The chaotic behaviour of the anisotropic Kepler problems was investigated in numerous
papers, see e.g. \cite{Casasayas::84,Devaney::82,Gutzwiller::90} and the non-integrability of planar problem was proved  in \cite{Gutzwiller::77}  and for planar and spatial problem in \cite{Arribas::03}.  The non-integrability proof in \cite{Arribas::03} uses the differential Galois approach and authors state that for $\mu\not\in\{0,1\}$ there is no  meromorphic integrals besides the Hamiltonian itself. But there is no written about meromorphic functions of what variables authors say. If one consider meromorphic functions of coordinates and momenta, then already Hamiltonian is not meromorphic function, thus system trivially is not meromorphically integrable for all values of $\mu$. Thus below we formulate these theorems in a more precise way. 

\begin{theorem}
 Hamiltonian system defined by~\eqref{eq:h2} is integrable in the Liouville sense with first integrals which are meromorphic in $(x,y,p_1,p_2,r)$ where
$r=\sqrt{x^2+\mu y^2}$, if and only if $\mu\in\{0,1\}$.
\end{theorem}

In the case when $\mu=1$ 
this system has two additional
functionally independent additional first integrals
\[
I_1=p_2x-p_1y,\qquad I_2=p_2 ( p_1 y-p_2 x)+\frac{ x}{\sqrt{x^2
    + y^2}},
\]
thus it is super-integrable.

Spatial anisotropic Kepler problem defined by \eqref{eq:h3}   has an invariant subspace defined by $z = p_3 = 0$. In this
subspace it coincides with the previous system. Thus, the necessary
conditions of the integrability are the same as for the previous system.
\begin{theorem}
  Hamiltonian system defined by~\eqref{eq:h3} is integrable in the Liouville sense with first integrals which are meromorphic in $(x,y,z, p_1,p_2,p_3,r)$, where \\ $r=\sqrt{x^2+\mu (y^2+z^2)}$, if and only if, $\mu\in\{0,1\}$.   
\end{theorem}

In the case $\mu=1$ it  coincides with three dimensional standard Kepler problem, and it has the following first integrals 
\[
\boldsymbol{c}=\vr  \times\vp, \qquad  \ve = \vp\times \vc  - \frac{\vr}{r},
\]
where 
$\vr=(x,y,z)$, $\vp=(p_1,p_2,p_3)$, and $r=\sqrt{x^2+y^2+z^2}$. 
Among them one can find three functionally independent and pairwise commuting. 

Hamiltonian~\eqref{eq:h2} (and also \eqref{eq:h3})  because of presence of square root $r$ is not single-valued and meromorphic in coordinates and momenta. Thus, formally, in order to apply the differential Galois theory approach to  such a Hamiltonian system  we have to extend it to the corresponding Poisson system introducing $r$ as additional variable. However, in calculations one can work with the original Hamiltonian system, and the only trace of this extension is the fact that we study the integrability in the class of meromorphic functions of not only coordinates and momenta but also of $r$. This extension procedure as well as its application to a certain  three-body problem was given in \cite{Maciejewski::13}. The similar trick is applied to all remaining Hamiltonian systems with algebraic potentials considered in this paper.

The above examples show that it is reasonable to examine similar
classes of constrained $n$-body systems.  In the next section we will
give several examples of such systems with a few degrees of freedoms.
In a case when the considered system reduces to a system with two
degrees of freedom the Poincar\'e cross sections give us quickly
insight into the dynamics of the systems.  However, a challenging problem 
is to prove that they are non-integrable and to find values of parameters for that they become integrable. For some presented problems we prove their non-integrability  using  the so-called \textit{Morales-Ramis} theory \cite{Morales::99}. It is based on analysis of differential Galois group of
variational equations obtained by linearisation of equations of motion
along a particular solution. The main theorem of this
theory states that if the considered system is meromorphically integrable in the
Liouville sense, then the identity component of the differential
Galois group of the variational equations is Abelian.  For a precise
definition of the differential Galois group and differential Galois
theory, see, e.g., \cite{van::03}.

\section{Integrability analysis of several restricted n-body problems}
\label{sec:2}
\subsection*{Model 3: Two masses on two inclined straight
  lines \label{sec:inclines}}
\begin{wrapfigure}{r}{0.38\textwidth}
  \vspace{-40pt}
  \begin{center}
    \includegraphics[width=0.28\textwidth]{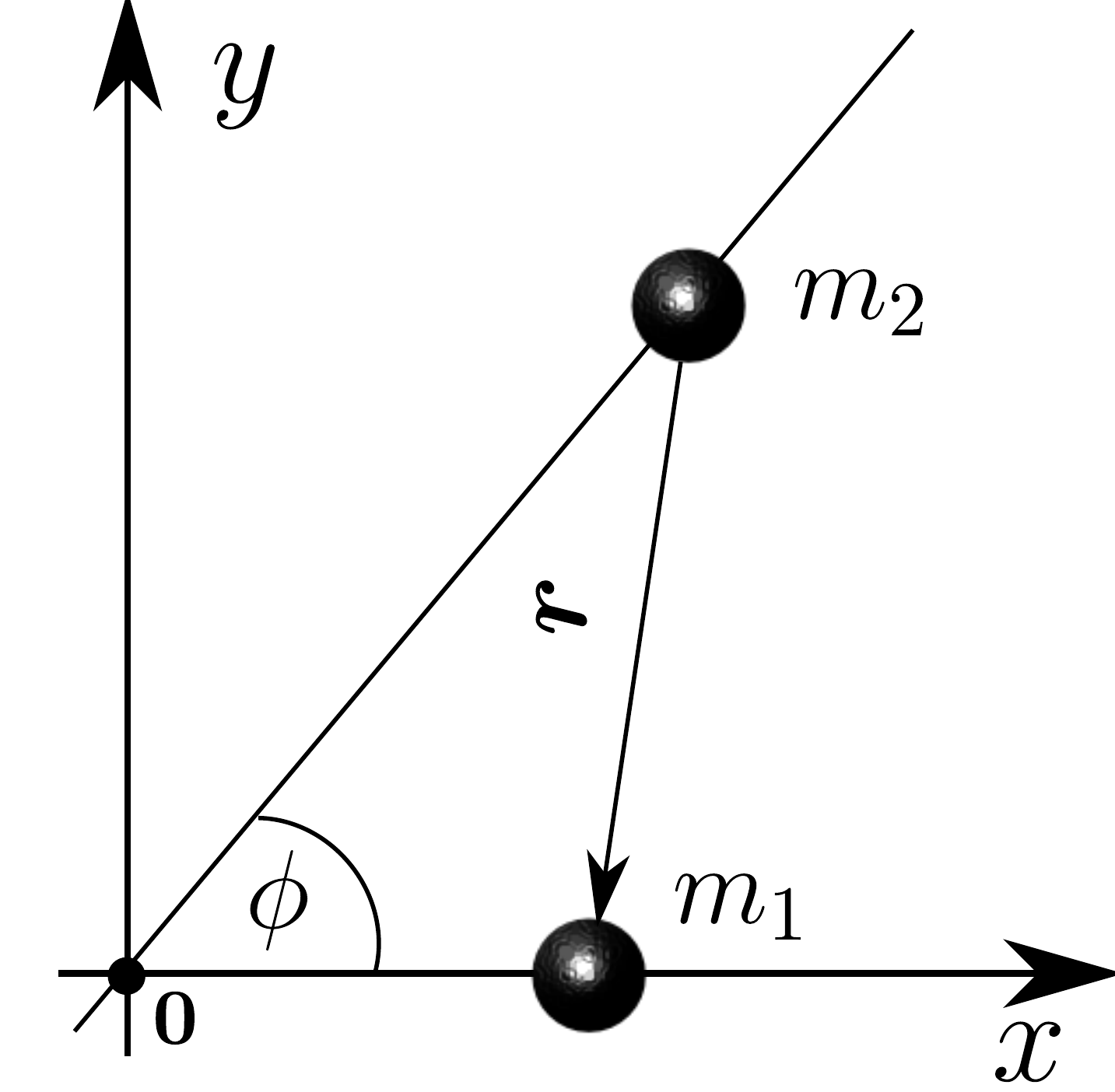}
  \end{center}
  \vspace{-15pt}
  \caption{Geometry of model 3. \label{fig:inclines}}
  \vspace{-25pt}
\end{wrapfigure}
The direct generalisation of the model 1 from Fig.\ref{fig:intro(a)}
is following.  Assume that mass $m_1$ moves along horizontal line
$q_2=0$ and it has coordinates $(q_1,0)$,  and mass   $m_2$ with coordinates $q_2(\cos\phi,\sin\phi)$ moves along a straight line inclined to the horizontal 
line, see Fig.\ref{fig:inclines}.  The Hamiltonian
function is given by
\begin{equation}
\label{eq:i}
  H=\frac{p_1^2}{2m_1}+\frac{p_2^2}{2m_2}-\frac{ G m_1 m_2}{\sqrt{q_1^2+q_2^2-2
      \cos\phi q_1 q_2}}.
\end{equation}
In Appendix we will prove the following
theorem.
\begin{theorem}
\label{thm:inclined}
  The system governed by Hamiltonian~\eqref{eq:i} is  integrable in the class of functions meromorphic 
  in $(q_1,q_2, p_1, p_2,r)$ where $r=\sqrt{q_1^2+q_2^2 -2q_1q_2 \cos \phi}$, iff 
  \begin{itemize}
  \item either $\phi\in\{0,\pi\}$
    and  $m_1, m_2\in \mathbb{R} $, or \\
  \item  $\phi\in\{\pi/2, 3\pi/2 \}$     and $m_2=m_1$.
  \end{itemize}
\end{theorem}
\subsection*{Model 4:  Masses moving on the parallel
  lines \label{sec:par}} \begin{wrapfigure}{r}{0.45\textwidth}
  \vspace{-0.45cm}
  \begin{center}
    \includegraphics[width=0.39\textwidth]{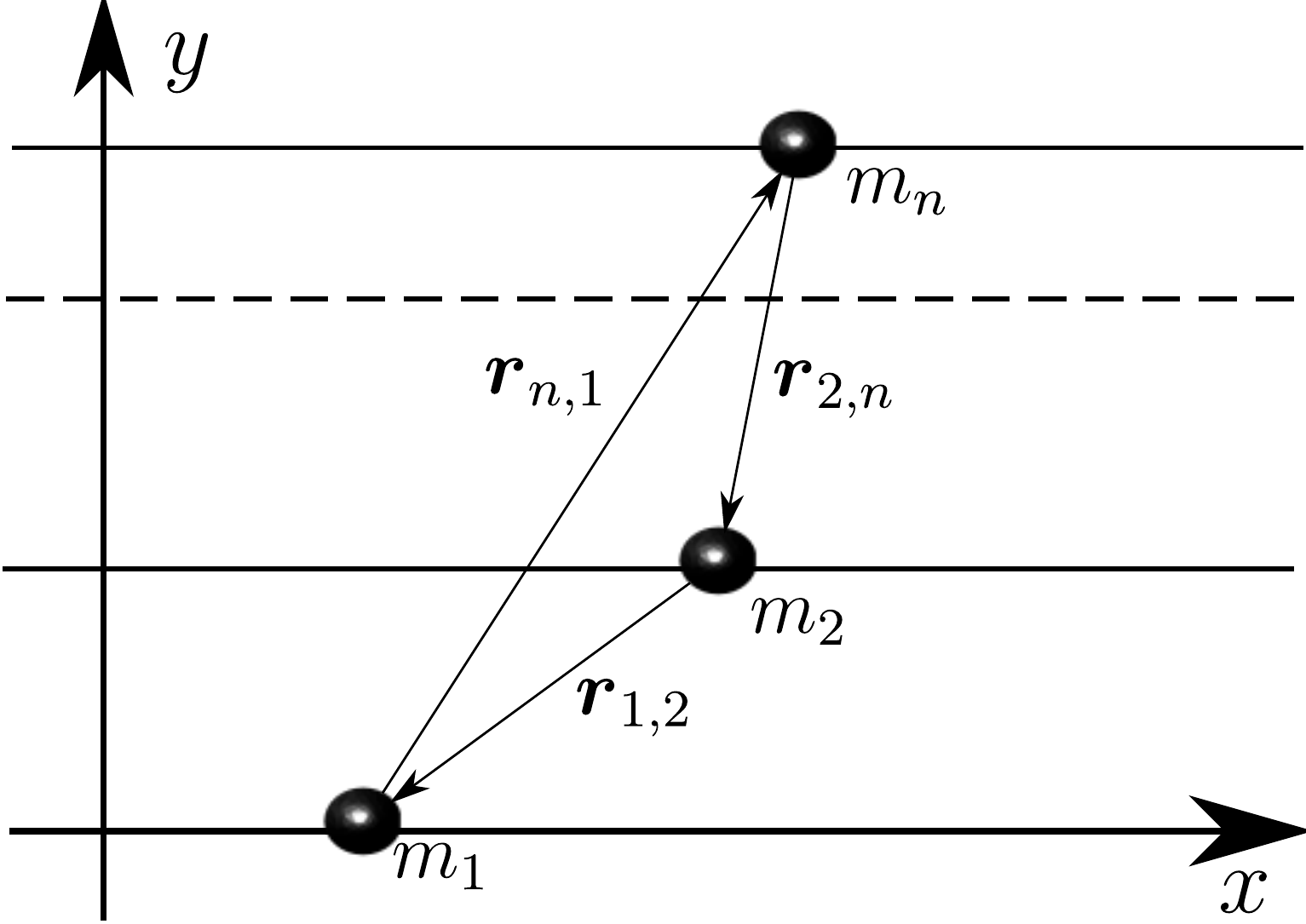}
  \end{center}
  \vspace{-15pt}
  \caption{Geometry of  model 4. \label{fig:par}}
  \vspace{-15pt}
\end{wrapfigure}
Let us consider a problem of $n$ masses moving in  parallel lines,
see Fig.~\ref{fig:par}.  As a generalised coordinates we use the
relative displacements $q_i=x_i-x_{i-1}$ along axis $x$, for $i=2,\ldots,n$ and $q_1=x_1$.

\begin{figure}[http]
  \centering \subfigure[$E=-2.25,$ ]{
    \includegraphics[width=0.47\textwidth]{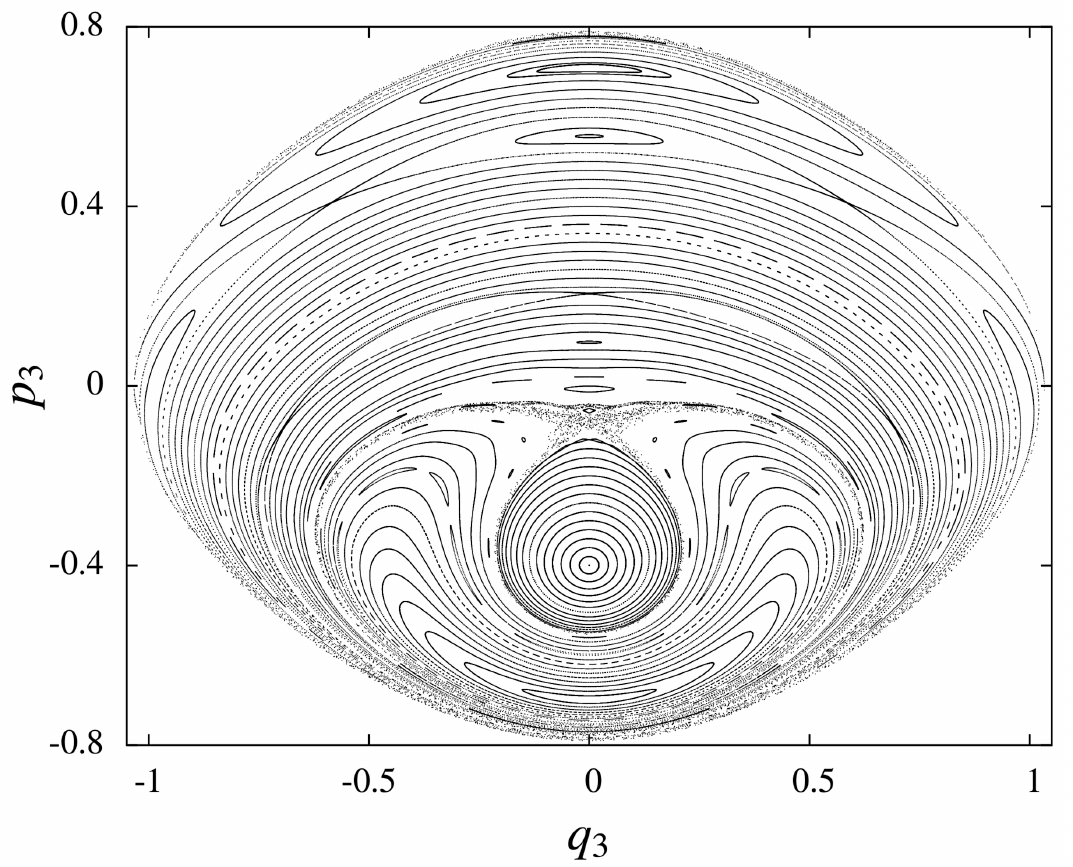}
  } \subfigure [$E=-2.1.$]{
    \includegraphics[width=0.47\textwidth]{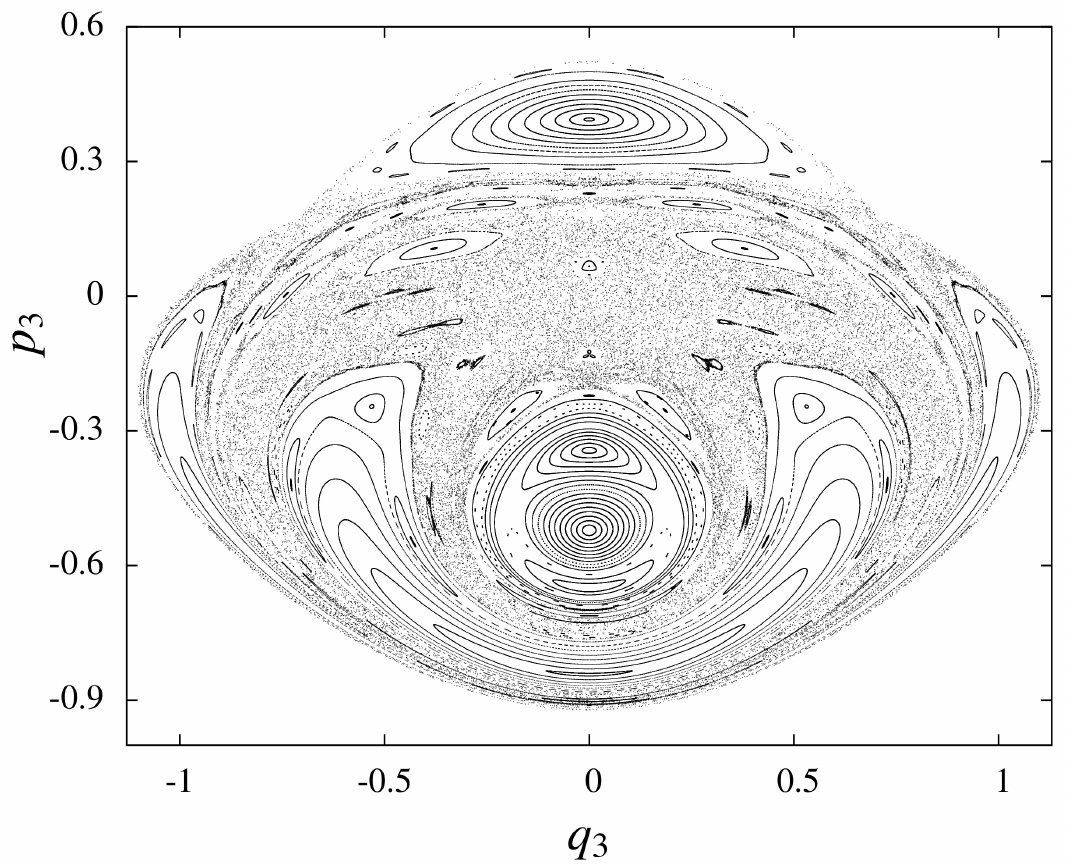}

  }
  \caption{Poincar\'e sections for model 4.  Parameters: $m_1=1,\
    m_2=2, \ m_3=1, \ G=1,$ \ $a=3,$ \newline $b=1, \ c=0,\ \text{cross-plain}
    \ q_2,\ p_2>0$. \label{poin:par}}
\end{figure}

The Lagrange and the
Hamiltonian functions do not depend on  variable $q_1$, which is a
cyclic variable and its corresponding momentum $p_1=c$ becomes a
parameter. Thus, we obtain the reduced system with $n-1$ degrees of
freedom. Model of $n=2$ masses is integrable. The reduced system with $n=3$
masses has two degrees of freedom and it is described by the following  Hamiltonian
\begin{equation}
  \label{eq:perp}
{\textstyle
    H=\frac{1}{2}
    \left(\frac{\left(c-p_2\right){}^2}{m_1}+\frac{\left(p_2-p_3\right){}^2}{m_2}+\frac{p
        _3^2}{m_3}-\frac{2 G m_2 m_3}{\sqrt{(a-b)^2+q_3^2}}       
         +m_1 \left(-\frac{2 G
          m_2}{\sqrt{a^2+q_2^2}}-\frac{2 G
          m_3}{\sqrt{b^2+\left(q_2+q_3\right){}^2}}\right)\right).}
\end{equation}
We assumed that masses $m_2$ and $m_3$ move along horizontal curves  $y=a$ and $y=b$, respectively. 
Fig.~\ref{poin:par} shows the Poincar\'e cross sections related to  \eqref{eq:perp}. Clearly, the system is generally non-integrable. However, a proof of this fact is an open question.  

\subsection*{Model 5: Two masses moving on an  ellipse and   a straight
  line parallel to the   main  axis of the ellipse \label{sec:elip+line} }
\begin{wrapfigure}{r}{0.38\textwidth}
  \vspace{-38pt}
  \begin{center}
    \includegraphics[width=0.38\textwidth]{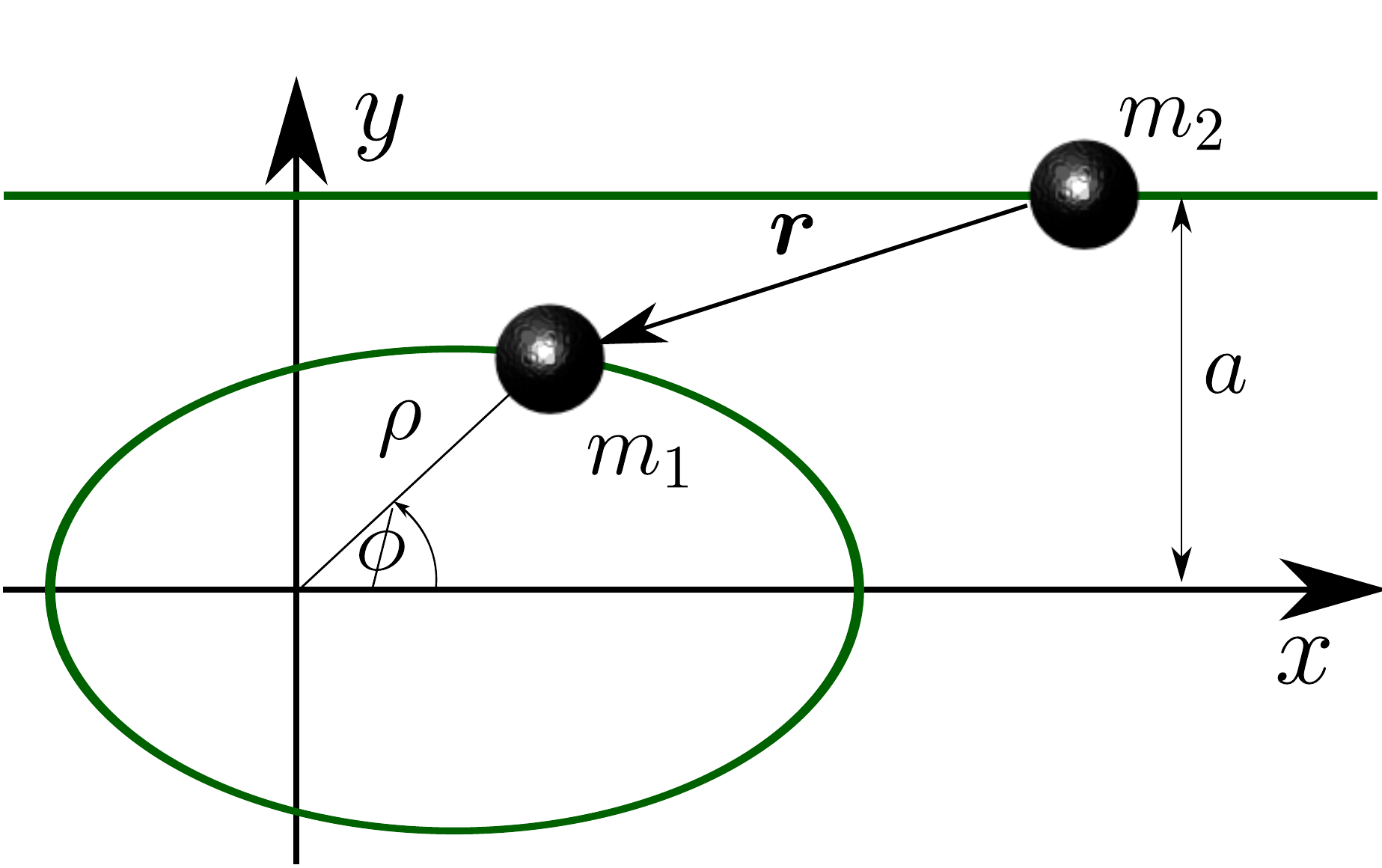}
  \end{center}
  \vspace{-10pt}
  \caption{Geometry of  model 5. \label{fig:elip+line}}
 \vspace{-25pt}
\end{wrapfigure}

In Fig.~\ref{fig:elip+line} the geometry of the system is shown.
 Now we assume that the mass $m_1$ moves on the ellipse with coordinates
$(\rho \cos\phi,\rho\sin\phi)$, where $ \rho=c/(1+e\cos\phi),$ and
mass $m_2$ moves along a straight line parallel to the main axis of
ellipse with coordinates $(x,a)$. The Hamiltonian function is given by
\begin{equation}
\label{eq:e}
 H=\frac{1}{2} \left(\frac{p_x^2}{m_2}+\frac{p_{\phi }^2 \left(1+e
        \cos\phi\right){}^4}{c^2 m_1 \left(1+e^2+2 e \cos\phi\right)}-\frac{2G m_1 m_2}{\sqrt{\left(\frac{c \cos\phi}{1+e
            \cos\phi}-x\right)^2+\left(\frac{c \sin\phi}{1+e
            \cos\phi}-a\right)^2}}\right).
\end{equation}

\begin{figure}[h!]
  \centering \subfigure[$E=-1,$]{
    \includegraphics[width=0.45\textwidth]{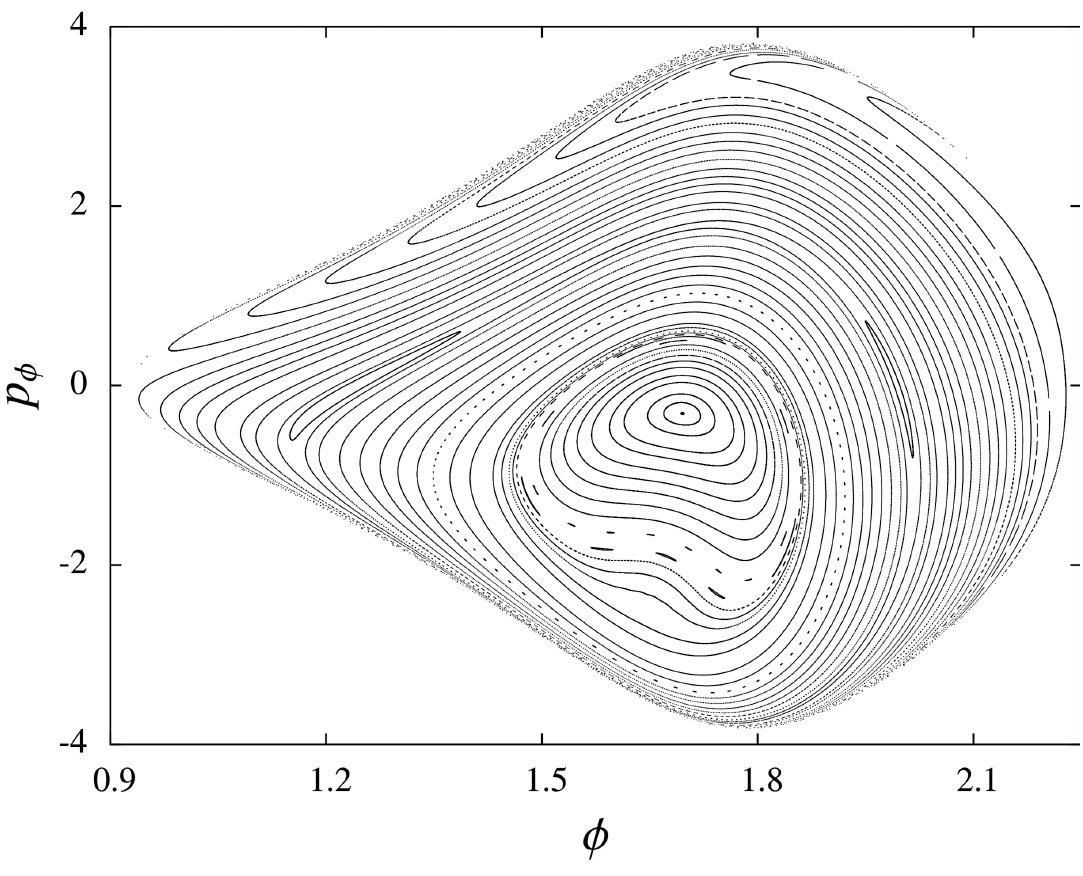}
  } \subfigure [$E=-0.5.$]{
    \includegraphics[width=0.45\textwidth]{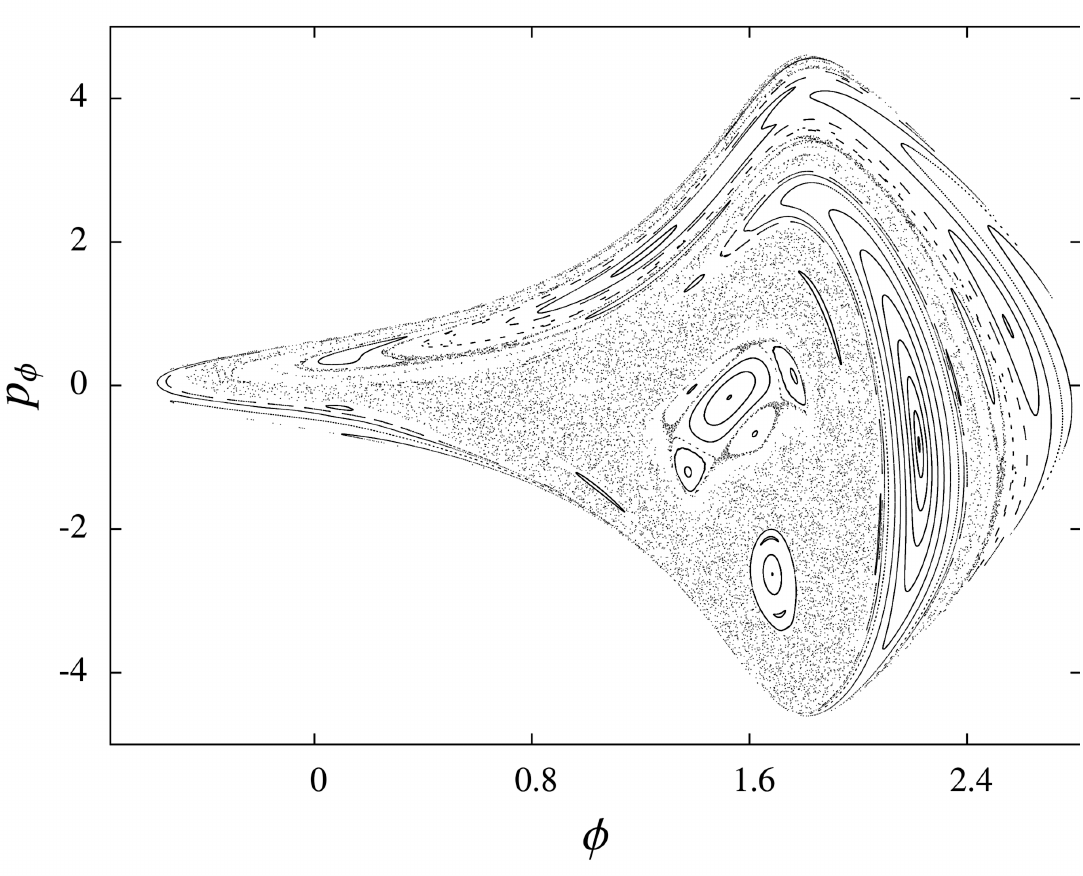}
  }
  \caption{ \label{poin:elip+line} Poincar\'e sections for model 5.
    Parameters: $m_1=1,\ m_2=2, \ G=1, \ a=3$, \ $c=2, \ e=0.5,$
    $\text{cross-plain} \ x,\ p_x>0$.}
\end{figure}

Fig.~\ref{poin:elip+line} shows the Poincar\'e cross sections. They present 
that for certain  fixed values of parameters, the system is 
not integrable. In fact  we can prove the following theorem. 
\begin{theorem}
If $a=0$, and  $m_1\neq m_2$, $m_1m_2\neq 0$, then the system governed by Hamiltonian~\eqref{eq:e} is not completely  integrable with first integrals  which are meromorphic in $(x,\phi, p_1, p_2,r)$, where
\[
r=\sqrt{\left(\frac{c \cos\phi}{1+e
            \cos\phi}-x\right)^2+\left(\frac{c \sin\phi}{1+e
            \cos\phi}\right)^2}.
\]

\end{theorem}
This theorem is in particular true for the circle when $e = 0$ and $c =
\rho$.
\subsection*{Model 6: Two mass points moving in two conics \label{sec:elipcic}}
\begin{figure}[h!]
  \centering \subfigure[Geometry of the model 6; ]{
    \includegraphics[width=0.45\textwidth]{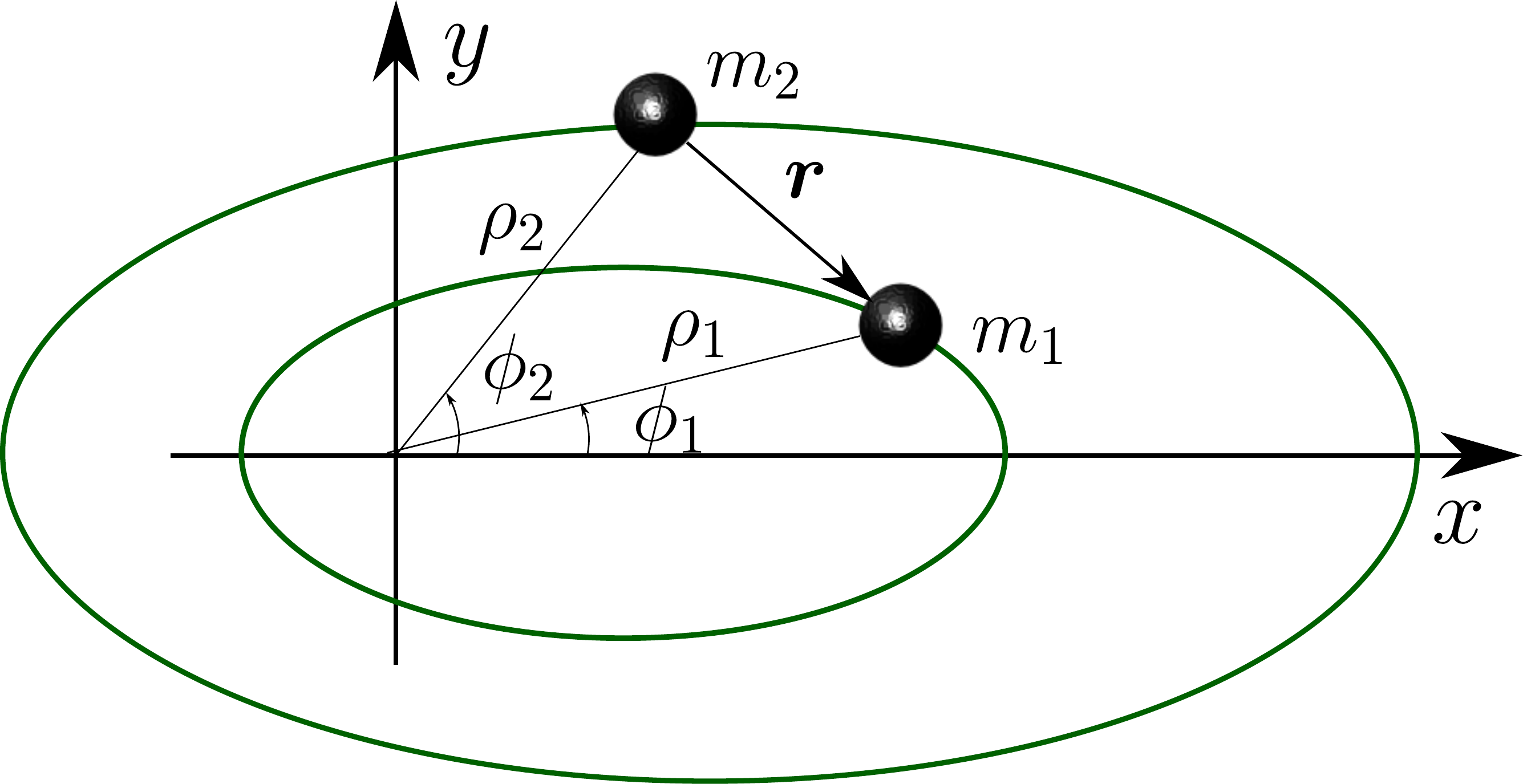}
    \hspace{0.2in} \label{fig:elipki(a)} } \subfigure[Geometry of the
  model 7.]{
    \includegraphics[width=0.45\textwidth]{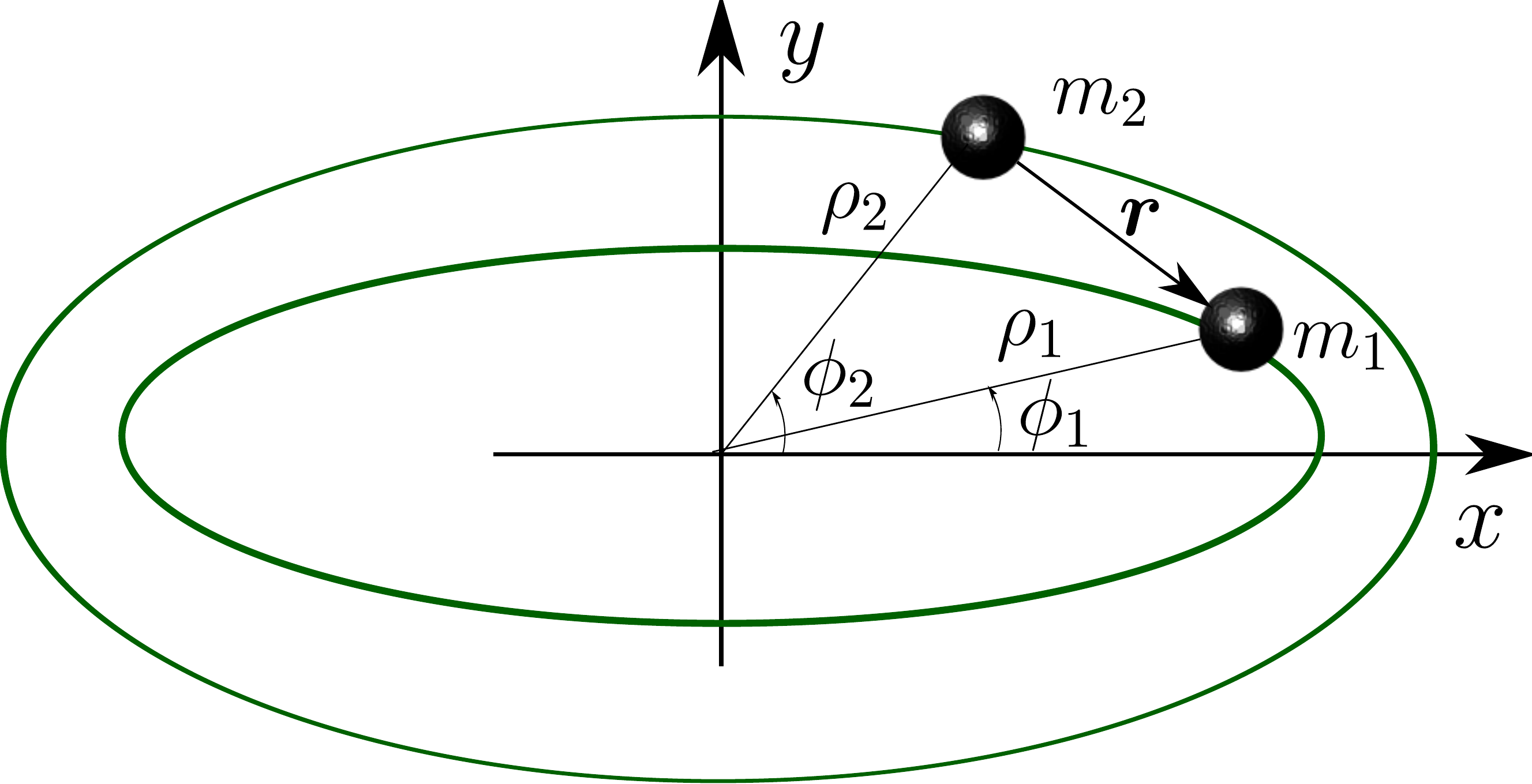}\label{fig:elipki(b)}
  }
  \caption{Motion of two masses on: (a) two confocal ellipses, and 
    (b) two concentric ellipses with parallel main axes.}
\end{figure}
\begin{figure}[http]
  \centering \subfigure[$E=-4.4,$]{
    \includegraphics[width=0.47\textwidth]{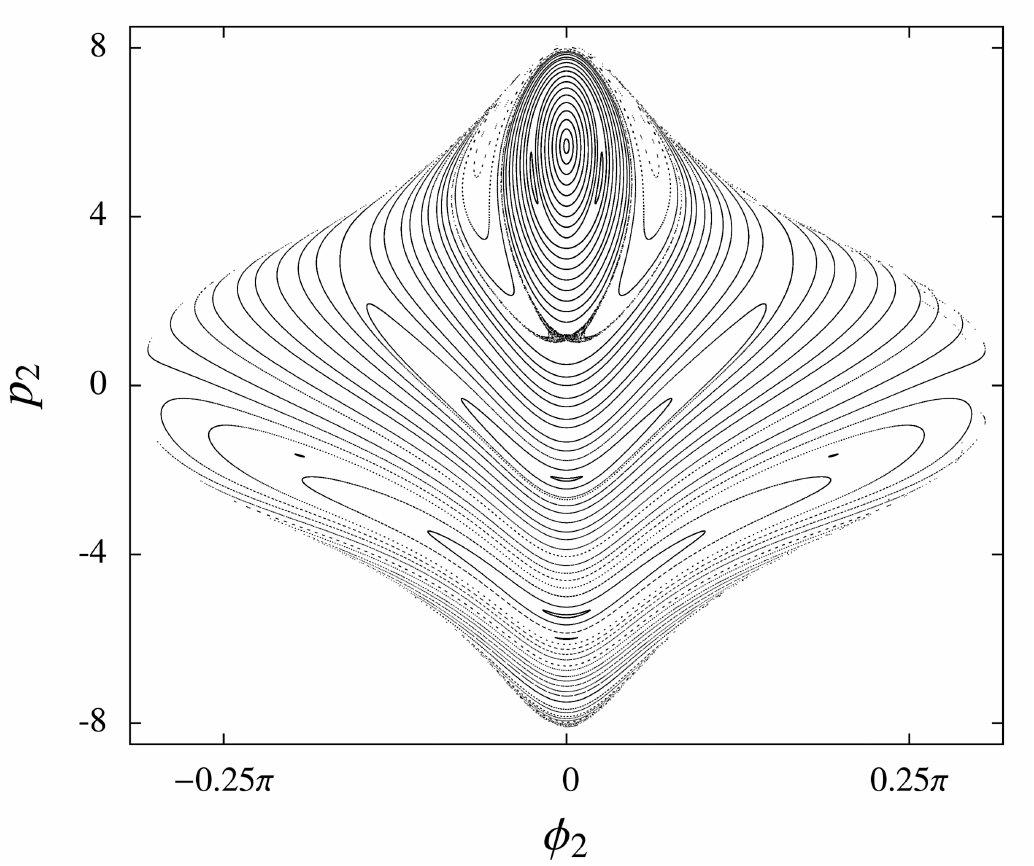}

  } \subfigure [$E=-1.8.$]{
    \includegraphics[width=0.47\textwidth]{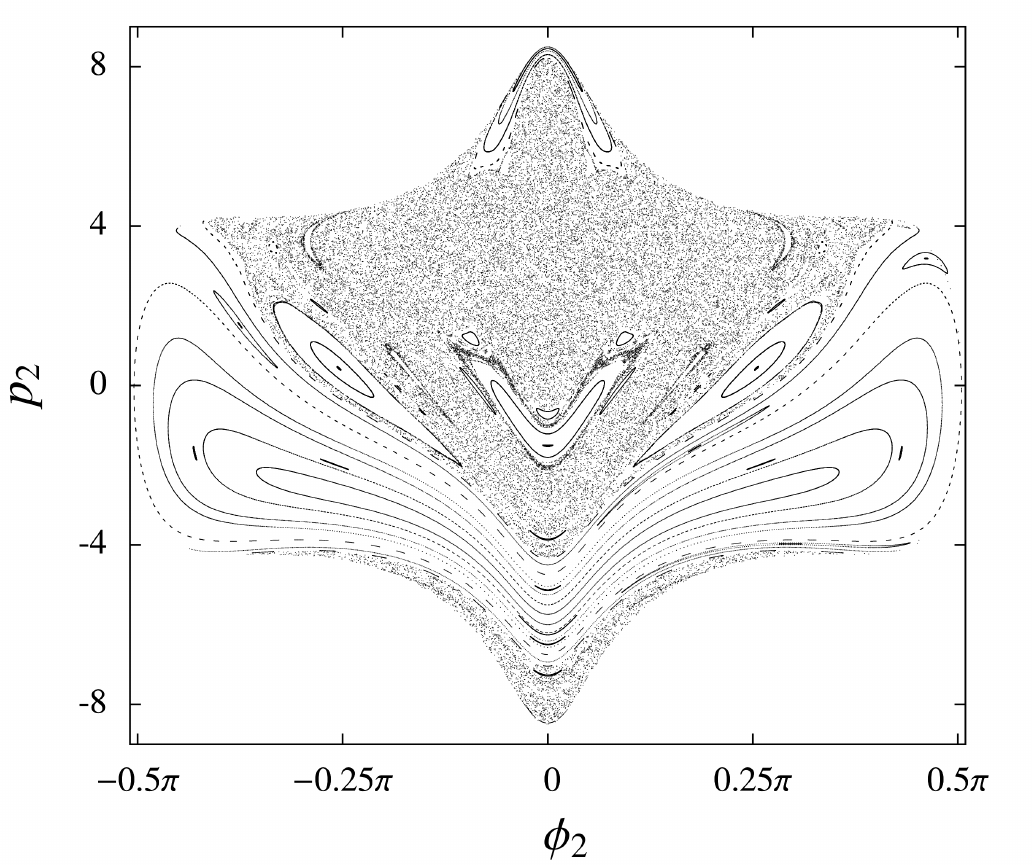}

  }
  \caption{Poincar\'e sections for model 6. Parameters: $m_1=2,\
    m_2=2, \ G=1, \ c_1=1,\ c_2=2,$ $e_1=\frac{1}{2},$
    $e_2=\frac{3}{2},$ $\text{cross-plain} \ \phi_1,\
    p_1>0$.\label{m6p}}
  \subfigure[$E=-0.7,$]{
    \includegraphics[width=0.47\textwidth]{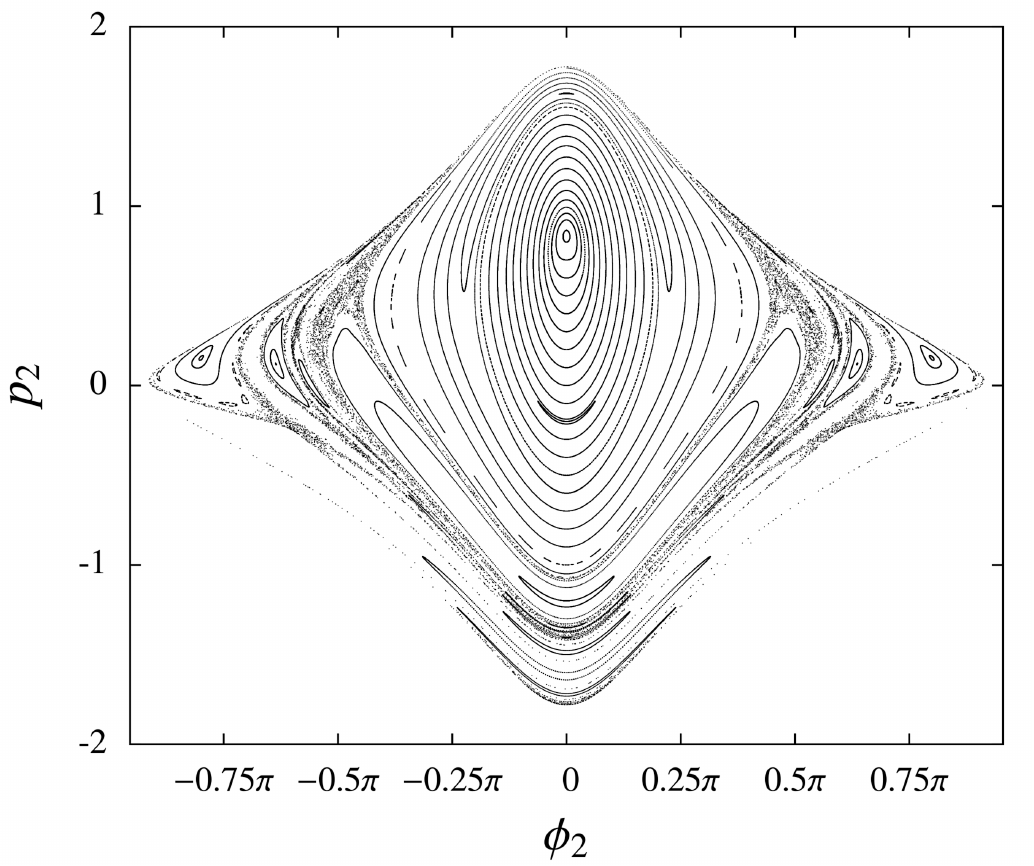}

  } \subfigure [$E=-0.5.$]{
    \includegraphics[width=0.47\textwidth]{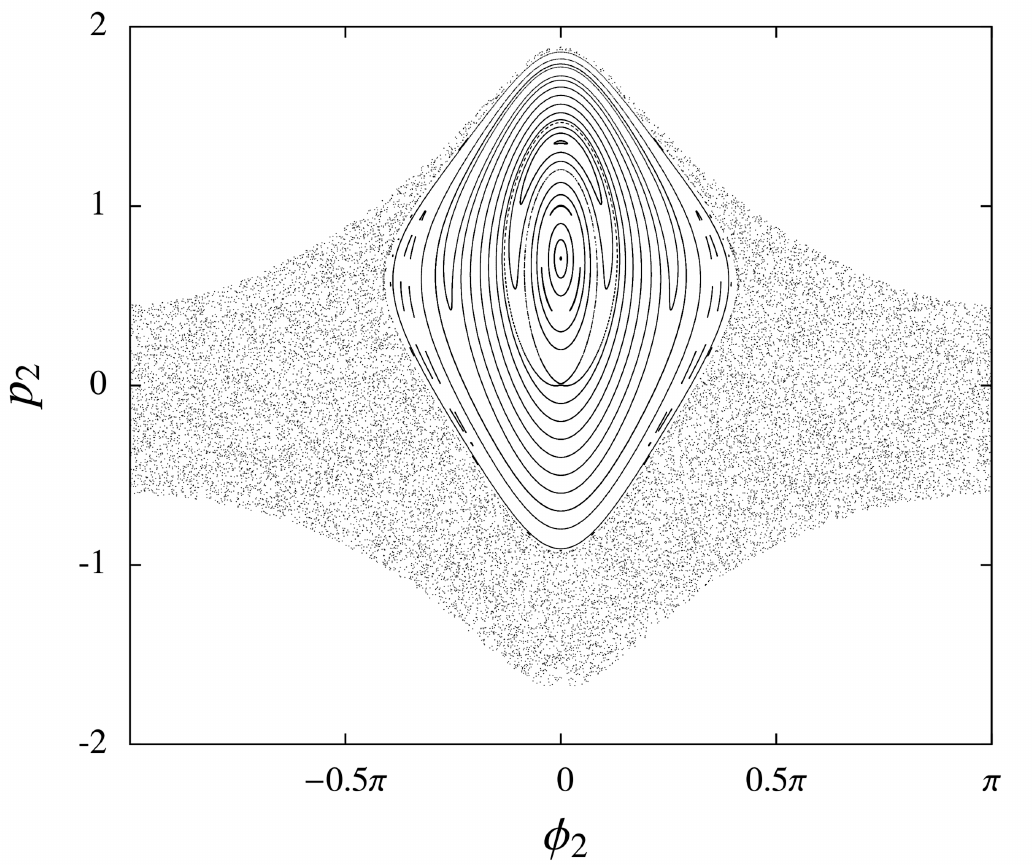}

  }
  \caption{Poincar\'e sections for model 6. Parameters: $m_1=2,\
    m_2=1, \ G=1, \ c_1=3,\ c_2=1,$ $e_1=\frac{3}{5},$ $e_2=0,$
    $\text{cross-plain} \ \phi_1,\ p_1>0$.\label{m7b}}
\end{figure}

In Fig.~\ref{fig:elipki(a)} the geometry of the system is
presented. In this case, masses $m_1$ and $m_2$ move along two
confocal ellipses with coordinates
$(\rho_1\cos\phi_1,\rho_1\sin\phi_1)$ and
$(\rho_2\cos\phi_2,\rho_2\sin\phi_2)$, where
\[
\rho_1=\frac{c_1}{1+e_1\cos\phi_1},\qquad
\rho_2=\frac{c_2}{1+e_2\cos\phi_2},
\]
and interact gravitationally. Hamiltonian function takes the form
\begin{equation}
  \begin{split}&
    H=\frac{
      \left(1+e_1 \cos \phi_1\right)^4p_1^2}{2 c_1^2 m_1 \left(2 e_1 \cos
        \phi_1+e_1^2+1\right)}+\frac{ \left(1+e_2 \cos \phi
        _2\right)^4p_2^2}{2c_2^2 m_2 \left(2 e_2 \cos\phi
        _2+e_2^2+1\right)}-\frac{G m_1 m_2}{B}, \\&
    B=\sqrt{\left(\frac{c_1 \cos \phi
          _1}{1+e_1 \cos \phi_1}-\frac{c_2 \cos \phi
          _2}{1+e_2 \cos \phi_2}\right)^2+\left(\frac{c_1 \sin
          \phi_1}{1+e_1 \cos \phi_1}-\frac{c_2 \sin
          \phi_2}{1+e_2 \cos \phi_2}\right)^2}.
  \end{split}
\end{equation}
To present the dynamics of considered system we make several
Poincar\'e cross sections, see Figs. \ref{m6p}-\ref{m7b}.

\subsection*{Model 7: Two masses moving  in concentric  ellipses with parallel main axes \label{sec:wspolelips}}
\begin{figure}[http]
  \centering \subfigure[$E=-0.82,$ ]{
    \includegraphics[width=0.47\textwidth]{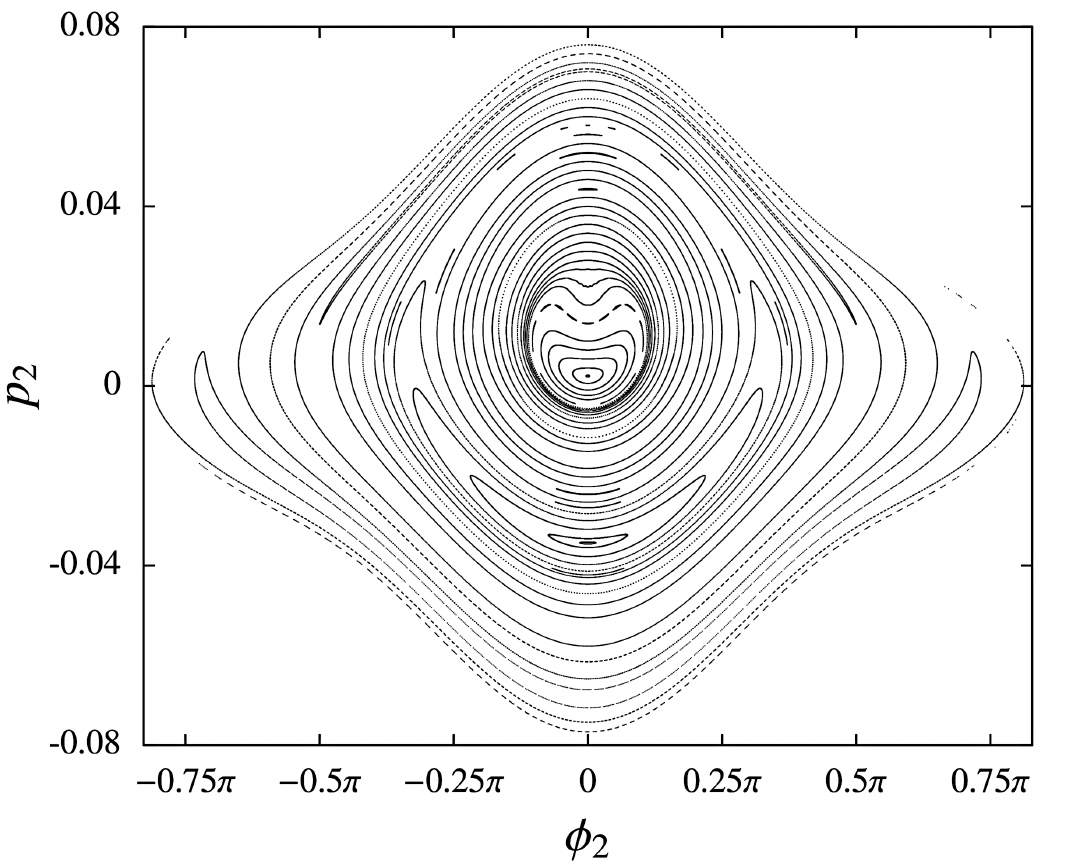}
  } \subfigure [$E=-0.8.$]{
    \includegraphics[width=0.47\textwidth]{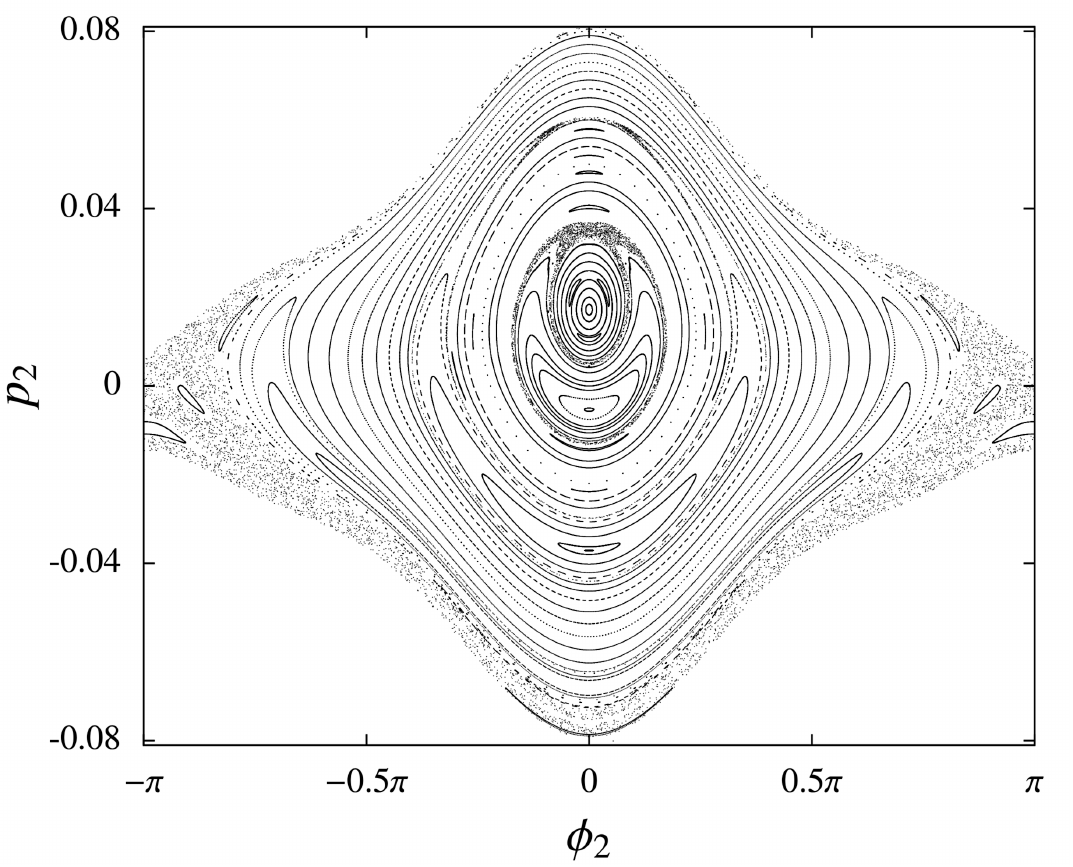}
  
  }
  \caption{Poincar\'e sections related to model 7.  Parameters:
    $m_1=1,\ G=1,$ \ $a_1=0.8,$ $\ a_2=1.1,$ $ \ b_1=1, \
    b_2=\frac{a_2b_1}{a_1}=1.4, \ m_2=\frac{m_1a_1}{a_2}= 0.73, $
    $\text{cross-plain} \ \phi_1,\ p_1>0$. \label{poin:wsw}}
\end{figure}

The geometry of the system is shown in Fig.~\ref{fig:elipki(b)}. In
this case, masses $m_1$ and $m_2$ move in two ellipses which have common centres and parallel main axes. Using the standard trigonometric parametrizations of points on ellipses $(a_i\cos\phi_i,b_i\sin\phi_i)$ for $i=1,2$, we can
derive the Hamiltonian
\begin{equation}
  \begin{split}
    & H=\frac{1}{2} \left(\frac{p_1^2}{a_1^2 m_1 \cos\phi_1^2+a_2^2
        m_1 \sin\phi_1^2}+\frac{p_2^2}{b_1^2 m_2 \cos\phi_2^2+b_2^2
        m_2 \sin\phi_2^2} \right. \\ - & \left. \frac{2 G m_1
        m_2}{\sqrt{\left(a_2 \cos\phi_1-b_2
            \cos\phi_2\right){}^2+\left(a_1 \sin\phi_1-b_1
            \sin\phi_2\right){}^2}}\right),
  \end{split}
\end{equation}
where $a_1, a_2$ and $b_1, b_2$ are major and minor
semi-axes. The Poincar\'e cross sections are shown  in
Fig.~\ref{poin:wsw}.

\subsection*{Model 8: N-masses moving in the
  circles \label{sec:3circles} }

   \begin{wrapfigure}{r}{0.45\textwidth}
       \vspace{-60pt}
       \begin{center}
         \includegraphics[width=0.42\textwidth]{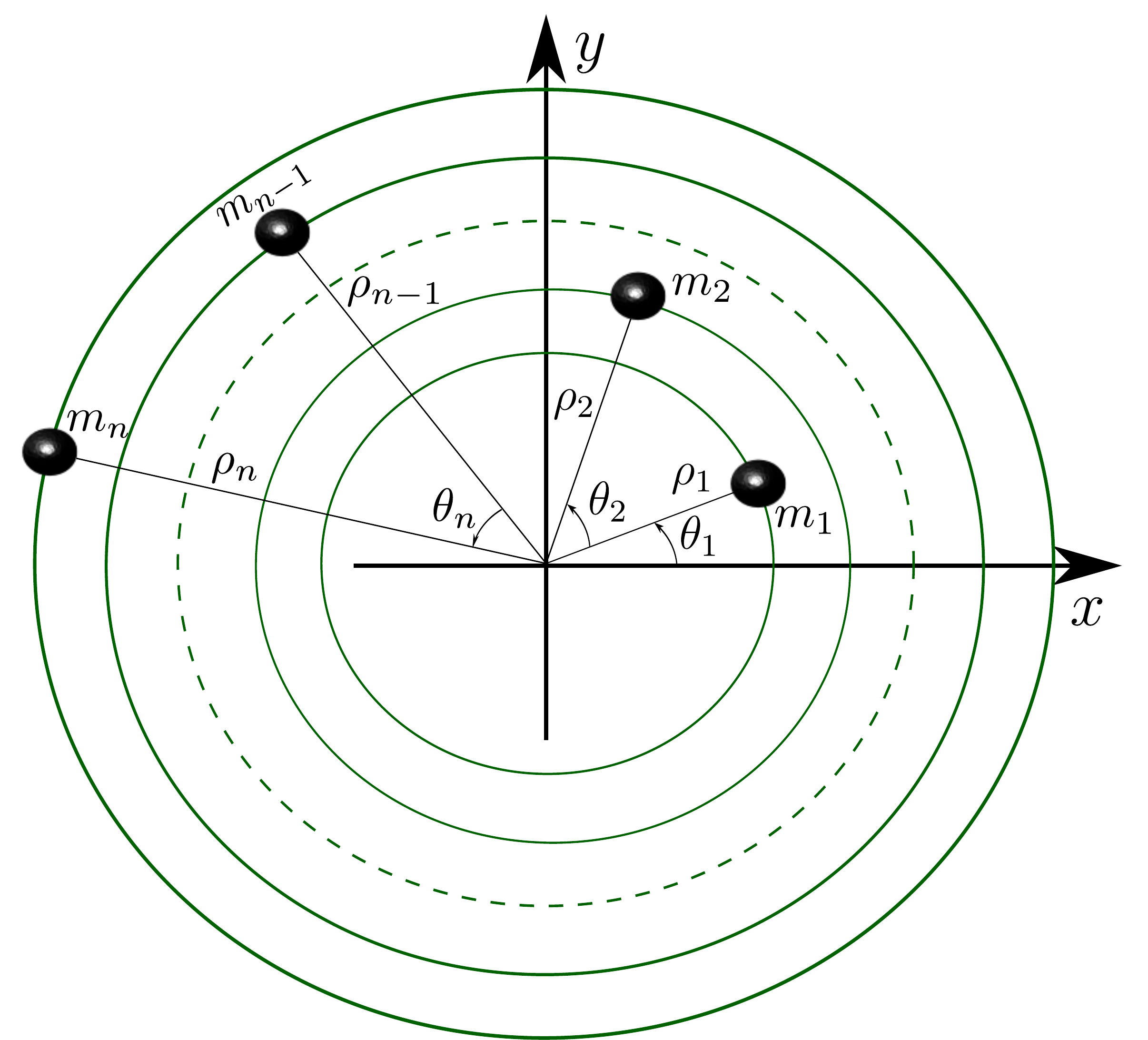}
       \end{center}
       \vspace{-15pt} 
       \caption{Geometry of  model 8. \label{fig:3circles}}
       \vspace{-15pt}
     \end{wrapfigure}
 
\begin{figure}[htt]
    \centering \subfigure[$E=-3.5,$ ]{
      \includegraphics[width=0.47\textwidth]{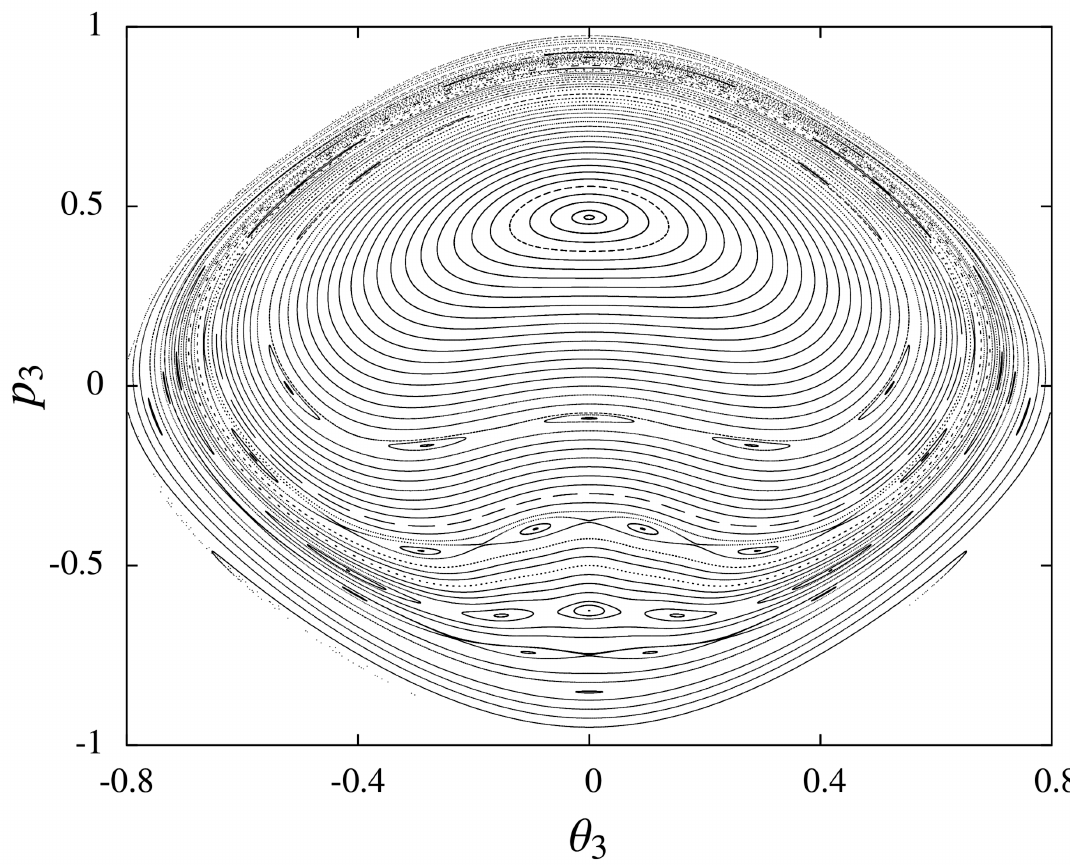}

    } \subfigure [$E=-2.8.$]{
      \includegraphics[width=0.47\textwidth]{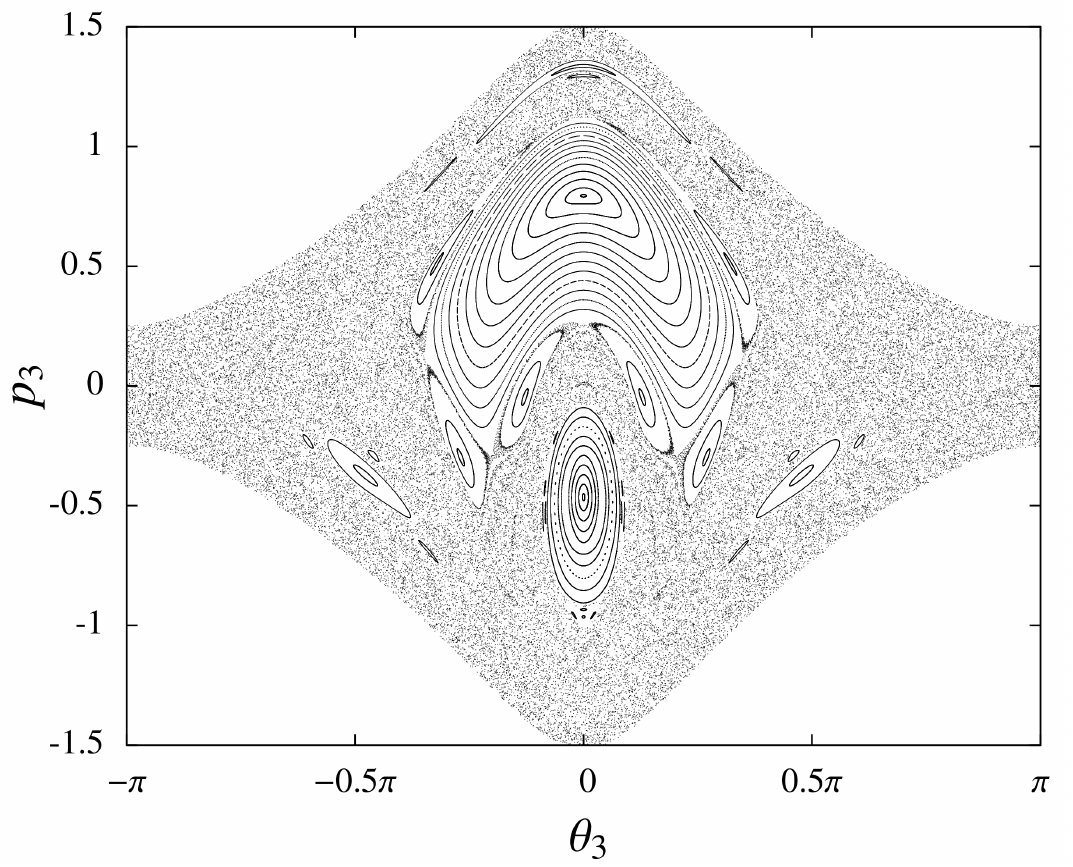}

    }
    \caption{Poincar\'e sections for model 8.  Parameters: $m_1=1,\
      m_2=2, \ m_3=1, \ G=1,$ \ $\rho_1=2, $ $ \rho_2=3, \ \rho_3=1, \
      c=0,$ $\text{cross-plain} \ \theta_2,\ p_2>0$. \label{poin:cic}}
  \end{figure}

     Let us consider the motion of $n$-masses moving on the concentric
     circles which interact gravitationally. As a generalized
     coordinates we use the relative angles $\theta_i$, see~Fig.~\ref{fig:3circles}. 
    Similarly to the fourth model 
     the Hamiltonian function has one
     cyclic variable $\theta_1$ and its corresponding momentum  $p_1=c$ is a first
     integral of the system. Thus, we get the reduced system with
     $n-1$ degrees of freedom. Case of two masses is of course
     integrable, but the model of $n=3$ has much more complex
     dynamics. To present this complexity we make several Poincar\'e
     sections, see Fig.~\ref{poin:cic}. Hamiltonian of this reduced system has the form
   \begin{equation}
       \begin{split} \label{eq:3cic}
         & H=\frac{1}{2} \left(\frac{\left(c-p_2\right){}^2}{\rho _1^2
             m_1}+\frac{\left(p_2-p_3\right){}^2}{\rho _2^2
             m_2}+\frac{p_3^2}{\rho _3^2 m_3}  -\frac{2 G m_1 m_2}{\sqrt{\rho
               _1^2+\rho _2^2-2 \rho _1 \rho _2 \cos\theta
               _2}} \right. \\ & \left. -\frac{2 G m_2 m_3}{\sqrt{\rho _2^2+\rho _3^2-2 \rho
               _2 \rho _3 \cos \theta _3}}-\frac{2 G m_1
             m_3}{\sqrt{\rho _1^2+\rho _3^2-2 \rho _1 \rho _3 \cos
               \left(\theta _2+\theta _3\right)}}\right).
       \end{split}
     \end{equation}

\section*{Acknowledgement}
The authors are very grateful to Andrzej J. Maciejewski for many helpful
comments and suggestions.
This research has been supported by grant No.~DEC-2011/02/A/ST1/00208 of
National Science Centre of Poland.
\section*{Appendix: Proof of Theorem~\ref{thm:inclined}}
\addcontentsline{toc}{section}{Appendix}
Hamilton equations for Hamiltonian \eqref{eq:i}  have the form
\begin{equation}
\label{eq:canonik}
\hspace*{-0.3cm}\dot{q}_1= \frac{p_1}{m_1},\ \dot{q}_2=\frac{p_2}{m_2},\
\dot{p}_1=\frac{G m_1 m_2
   \left(\alpha  q_2-q_1\right)}{\left(q_1^2+q_2^2-2 \alpha  q_1
   q_2\right){}^{3/2}},
  \ \dot{p}_2= \frac{G m_1 m_2
   \left(\alpha  q_1-q_2\right)}{\left(q_1^2+q_2^2-2 \alpha  q_1
   q_2\right){}^{3/2}},
\end{equation}
where $\alpha:= \cos\phi$.
In order to simplify calculations, we make the following non-canonical transformation
\begin{equation}
\begin{split}
&\begin{bmatrix}
q_1\\
q_2\\
p_1\\
p_2\\
\end{bmatrix}=
\left[
\begin{array}{cccc}
 \frac{1}{\sqrt{\beta ^2}-\mu_2} &
   \frac{\sqrt{\beta^2-\mu_2^2} (\mu_1+\mu_2)}{\left(\mu_2-\sqrt{\beta ^2}\right)
   \sqrt{\mu_1^2-\mu_2^2)}} & 0 & 0 \\
 0 & 1 & 0 & 0 \\
 0 & 0 & \frac{\mu_1-\mu_2}{2
   \sqrt{\beta ^2}-2 \mu_2} &
   \frac{\sqrt{\beta^2-\mu_2^2} \sqrt{\mu_1^2-\mu_2^2}}{\left(\mu_2-\sqrt{\beta ^2}\right)
   (\mu_1+\mu_2)} \\
 0 & 0 & 0 & 1 \\
\end{array}
\right]
\begin{bmatrix}
x_1\\
x_2\\
y_1\\
y_2\\ \end{bmatrix},\\
 &
\mu_1= m_1+m_2, \quad \mu_2=m_2-m_1, \quad \beta=\sqrt{\left(m_1-m_2\right)^2+4m_1m_2\alpha^2}.
\end{split}
\label{eq:noncal}
\end{equation}
System \eqref{eq:canonik} after this transformation takes the form
\begin{equation}
\label{eq:transcan}
\begin{split}
&\dot{x}_1=y_1, \,\,
\dot{y}_1=\frac{G x_1 \left(\mu _1-\beta \right) \left(\mu
   _2-\beta \right){}^3}{2 \left(-2 x_2 x_1 \left(\beta
   +\mu _1\right) \sqrt{\frac{\beta ^2-\mu _2^2}{\mu
   _1^2-\mu _2^2}}+\frac{2 \beta  x_2^2 \left(\beta +\mu
   _1\right) \left(\beta -\mu _2\right)}{\mu _1-\mu
   _2}+x_1^2\right){}^{3/2}}, \\ 
   &
\dot{x}_2=\frac{2y_2}{\mu_1+\mu_2},\,\, {\textstyle  \dot{y}_2=\frac{G \left(\beta -\mu _2\right){}^2 \left(x_1
   \sqrt{\left(\mu_1^2-\mu_2^2\right) \left(\beta^2-\mu_2^2\right)}-\left(\mu _1+\mu _2\right) x_2 \left(\beta
   +\mu _1\right) \left(\beta -\mu _2\right)\right)}{4
   \left(-2 x_2 x_1 \left(\beta +\mu _1\right)
   \sqrt{\frac{\beta ^2-\mu _2^2}{\mu _1^2-\mu
   _2^2}}+\frac{2 \beta  x_2^2 \left(\beta +\mu
   _1\right) \left(\beta -\mu _2\right)}{\mu _1-\mu
   _2}+x_1^2\right){}^{3/2}}}.
\end{split}
\end{equation}
It has invariant manifold
$
\scN=\left\{(x_1,x_2,y_1,y_2)\in\mathbb{C}^4\,|\,  x_1=y_1=0\right\}
$
and its restriction  to $\scN$  is 
\begin{equation}
\dot{x}_2=\frac{2y_2}{\mu_1+\mu_2}, \qquad \dot{y}_2= -\frac{G \left(\mu _1+\mu _2\right) \left[\left(\mu
   _1-\mu _2\right) \left(\beta -\mu
   _2\right)\right]{}^{3/2}}{8 \sqrt{2} x_2^2
   \sqrt{\beta ^3 \left(\beta +\mu _1\right)}}.
   \label{eq:parsol}
\end{equation}
Let  be the particular solution of \eqref{eq:transcan} defined by \eqref{eq:parsol}, and $\vZ=\left[X_1,X_2,Y_1,Y_2\right]^T$ denotes the variations of $[x_1,x_2,y_1,y_2]^T$. Then,  the variational equations along this particular solution have the form $\dot\vZ=A\vZ$, where
\[
A=
\left[
\begin{array}{cccc}
 0 & 0 & 1 & 0 \\
 0 & 0 & 0 & \frac{2}{\mu _1+\mu _2} \\
 \frac{G \left(\mu _1-\beta \right) \left(\mu _1-\mu
   _2\right){}^{3/2} \left(\mu _2-\beta \right){}^3}{4
   \sqrt{2} \beta ^{3/2} x_2^3 \left(\beta +\mu
   _1\right){}^{3/2} \left(\beta -\mu _2\right){}^{3/2}}
   & 0 & 0 & 0 \\
 -\frac{G \left(\beta +3 \mu _1\right) \left(\beta -\mu
   _2\right) \left(\mu _1-\mu _2\right){}^2 \sqrt{\beta
   +\mu _2} \sqrt{\mu _1+\mu _2}}{16 \sqrt{2} \beta
   ^{5/2} x_2^3 \left(\beta +\mu _1\right){}^{3/2}} &
   \frac{G \left(\beta -\mu _2\right){}^{3/2} \left(\mu
   _1-\mu _2\right){}^{3/2} \left(\mu _1+\mu
   _2\right)}{4 \sqrt{2} \beta ^{3/2} x_2^3 \sqrt{\beta
   +\mu _1}} & 0 & 0 \\
\end{array}
\right].
\]
Equations for $X_1$ and $Y_1$ form a subsystem of normal variational equations and can be rewritten as a one second-order differential equation for variable $X\equiv X_1$
\begin{equation}
\ddot{X}+ \left(-\frac{G \left(\mu _1-\mu _2\right){}^{3/2} \left(\beta
   -\mu _1\right) \left(\beta -\mu _2\right){}^{3/2}}{4
   \sqrt{2} \beta ^{3/2} x_2^3 \left(\beta +\mu
   _1\right){}^{3/2}}\right)X=0.
   \label{eq:onrma}
\end{equation}
We transform this equation using the following change of independent variable
\begin{equation}
t\longrightarrow z= -\frac{4 \sqrt{2} E\sqrt{\frac{\beta  \left(\beta +\mu
   _1\right)}{\left(\mu _1-\mu _2\right) \left(\beta
   -\mu _2\right)}}}{G \left(\mu _1^2- \mu _2^2\right)} x_2(t),
\end{equation}
where $E$ is a level of Hamiltonian transformed by means of \eqref{eq:noncal} and restricted to $\scN$.
Then normal variational equation~\eqref{eq:onrma} takes the form
\begin{equation}
X'' + pX' + qX=0,  \quad 
p=-\frac{1}{2z}+\frac{1}{2(z-1)}, \quad q=\frac{-\beta+\mu_1}{2(\beta+\mu_1)z^2 }+\frac{\beta-\mu_1}{2(\beta+\mu_1)(z-1)z},
\end{equation}
where $ '\equiv\Dz$.
We recognize  that this equation is a Riemann $P$ equation, see e.g., \cite{Kimura::70, Morales::99}
\begin{equation} 
\dfrac{\mathrm{d}^2X}{\mathrm{d}z^2}+\left(\dfrac{1-a-a'}{z}+ 
\dfrac{1-c-c'}{z-1}\right)\dfrac{\mathrm{d}X}{\mathrm{d}z}+ 
\left(\dfrac{aa'}{z^2}+\dfrac{cc'}{(z-1)^2}+ 
\dfrac{bb'-aa'-cc'}{z(z-1)}\right)X=0, 
\label{eq:riemann}
\end{equation}
with exponents
\begin{equation}
{\textstyle
a=\frac{1}{4}\left(3+\sqrt{1+\frac{16\beta}{\beta +\mu_1}}\right), \, a'=\frac{1}{4}\left(3-\sqrt{1+\frac{16\beta}{\beta +\mu_1}}\right), \,
b=c'=0, \,\,\, b'=-1, \,\,\, c=\frac{1}{2}.}
\label{eq:diffexpr1}
\end{equation}
The differences of exponents are given by
\begin{equation}
\lambda=a-a'=\frac{1}{2}\sqrt{\frac{17\gamma+1}{\gamma+1}}, \qquad\sigma=b-b'=1, \qquad \nu=c-c'=\frac{1}{2},
\end{equation}
where $\gamma=\beta/\mu_1$. Riemann $P$ equation is solvable iff one of the four numbers $\lambda+\sigma+\nu$,
  $-\lambda+\sigma+\nu$, $\lambda-\sigma+\nu$, $\lambda+\sigma-\nu$ is an odd
  integer or $\lambda$ or $-\lambda$ and $\sigma$ or $-\sigma$ and
  $\nu$ or $-\nu$ belong (in an arbitrary order) to the so-called Schwarz's table \cite{Kimura::70, Morales::99}.
Conditions $\pm\lambda+\sigma+\nu=2p+1$, where $p\in\Z$, give the following expression for $\gamma$
\begin{equation*}
\gamma=\frac{-3 + 5 p - 2 p^2}{1 - 5 p + 2 p^2},
\end{equation*}
that takes only two non-negative values 0 and 1. Similarly, conditions $\lambda-\sigma+\nu=2p+1$, and  $\lambda+\sigma-\nu=2p-1$, where $p\in\Z$, give
\[
\gamma=\frac{p - 2 p^2}{-2 - p + 2 p^2},\quad \gamma=\frac{-1 + 3 p - 2 p^2}{-1 - 3 p + 2 p^2},
\]
respectively, that only take two non-negative values 0 and 1.

Since two differences of exponents are equal to 1/2 and 1, only the first case in the Schwarz's table is admissible that leads to the condition $\lambda=1/2+p$, where $p\in\Z$.  It gives
\[
\gamma=-\frac{p + p^2}{-4 + p + p^2},
\]
and this expression takes only two non-negative values 0 and 1. Value $\gamma=1$ gives $\alpha=\cos\phi=\pm 1$, and that implies $\phi\in\{0,\pi\}$. Parameter $\gamma$ vanishes only when $\beta=0$, that gives $m_2=m_1$, and simultaneously $\alpha=\cos\phi=0$.  These are the only cases when the identity component of differential Galois group of  Riemann $P$ equation~\eqref{eq:riemann} with exponents~\eqref{eq:diffexpr1} is solvable that is necessary  for Abelianity and the integrability  of the system..

%
%
%

\end{document}